# Toward Scalable Machine Learning and Data Mining: the Bioinformatics Case

Faraz Faghri[1,2], Sayed Hadi Hashemi[1], Mohammad Babaeizadeh[1], Mike A. Nalls[2], Saurabh Sinha[1], Roy H. Campbell[1]

[1]Department of Computer Science, University of Illinois at Urbana-Champaign, Urbana, IL, USA
[2]Laboratory of Neurogenetics, National Institute on Aging, National Institutes of Health, Bethesda, MD, USA

**Abstract**

In an effort to overcome the data deluge in computational biology and bioinformatics and to facilitate bioinformatics research in the era of big data, we identify some of the most influential algorithms that have been widely used in the bioinformatics community. These top data mining and machine learning algorithms cover classification, clustering, regression, graphical model-based learning, and dimensionality reduction. The goal of this study is to guide the focus of scalable computing experts in the endeavor of applying new storage and scalable computation designs to bioinformatics algorithms that merit their attention most, following the engineering maxim of "optimize the common case".

**Introduction**

Biological data and particularly genomics data pose some of the most severe computational challenges facing us in the next decade [2]. Considering the computational demands across the lifecycle of a dataset—acquisition, storage, distribution, and analysis—genomics is either on par with or the most demanding of the big data domains. To address the storage and analysis challenges posed by big data, analytical systems are transitioning from shared, centralized architectures to distributed, decentralized architectures. Frameworks such as MapReduce, and its open-source variant Hadoop, exemplify this effort by introducing a programming model to facilitate efficient, distributed algorithm execution while abstracting away lower-level details [11]. However, these distributed frameworks are not suited for all purposes, and in many cases can even result in poor performance [14]. Algorithms that make use of multiple iterations, especially those using graph or matrix data representations, are particularly poorly suited for popular big data processing systems. For instance, machine learning and data mining involve iterative algorithms that are notoriously difficult to scale and parallelize, often due to inherent interdependencies within the computation steps and also the training data [15]. There are four key challenges with iterative machine learning and data mining algorithms in a distributed environment: first, training data is distributed on multiple nodes and the algorithms have to be parallelized in a way that each machine can be assigned an independent task; second, designing a termination condition which detects when a fix point has been reached; third, guaranteeing the consistency, convergence, and reproducibility to a relevant result; fourth, intelligently reducing the network overhead while efficiently utilizing I/O, memory, and CPU resources. In response to these challenges, new frameworks including Apache Mahout [12], Spark MLlib [13], and more recently Google's TensorFlow [25], and Microsoft's CNTK [26] have been developed with the potential to transform the ways in which researchers solve certain machine learning problems.

However, these computing frameworks have focused on algorithms which are widely used for social network analysis and advertisement, serving the needs of major tech, marketing, and finance industries. While these algorithms are providing satisfactory results in their respective areas, they do not necessarily provide the same performance in other fields, particularly bioinformatics. The goal of this study is to guide the focus of scalable computing experts in the endeavor of applying new storage and scalable computation designs to bioinformatics algorithms that merit their attention most, following the engineering maxim of "optimize the common case".

## Top Machine Learning and Data Mining Algorithms

A number of Machine Learning and Data Mining (MLDM) methods are commonly used by bioinformaticians today. The bioinformatics field has made substantial progress through use of these algorithms, but by and large, there has not been a systematic survey of the most popular algorithms in the bioinformatics community. We are interested in this question from the perspective of computer scientists looking to identify algorithms where to invest the most effort for improved scalability and efficiency. Identifying the primary algorithms is a subjective task requiring us to define clear criteria for making comparisons. Popularity, usefulness, and novelty are some possible criteria. Here, as part of a systematic review, we choose to compare algorithms by the number of times they have been cited in the bioinformatics literature, and thus identify the MLDM algorithms most widely used by the bioinformatics community. These algorithms cover classification, clustering, regression, graphical model-based learning, and dimensionality reduction, which are among the most important topics in machine learning and data mining research and development.

As a first step, we compiled a list of 26 well known MLDM algorithms pertaining to the above-mentioned tasks (Supplementary Note 1). We then surveyed publications in a selection of reputable bioinformatics and computational biology journals according to Google Scholar metrics in late 2015 [1]; the selection included the top five journals by 'h5 index': Bioinformatics, BMC Bioinformatics, PLOS Computational Biology, Journal of Theoretical Biology, and BMC Systems Biology. Specifically, we counted the papers with at least one occurrence of each algorithm's name in full-text of articles published in these journals. Since the same algorithm may be referred to by different names in scientific literature, we queried for various possible names for each algorithm (Supplementary Note 1) and aggregated the results. This process identified studies that have used or implemented a specific algorithm, or refer to other relevant studies that do so. We undertook this systematic review using Google Scholar in late 2015, and removed those algorithms (from the initial compilation) that did not have at least 50 papers citing/using them. The remaining 20 algorithms were then organized into 5 topics: dimension reduction, graphical model-based learning, regression, clustering, and classification.

Figure 1 shows that Hidden Markov Models (HMM) and Support Vector Machines (SVMs) are two of the most commonly used algorithms in bioinformatics, by our criteria. HMMs are used most commonly in genome annotation tasks such as gene finding [3], enhancer prediction [4], protein domain annotation [5] prediction, haplotype phasing [6] and more recently in unsupervised learning of chromatin states [7]. SVMs have been employed in a wide array of classification tasks in bioinformatics, such as gene expression analysis [8], cancer classification [9], and protein subcellular localization prediction [10].

The top five algorithms include one classification method (SVM), a clustering method (Hierarchical Clustering), a graphical model method (HMM), linear regression for value prediction or testing an association and Principal Components Analysis (PCA) for dimensionality reduction and data exploration.

We noted that the observed trends are largely consistent across journals, although journal-specific biases towards one method over another were seen to an extent. For example, while overall ranked as third, linear regression is the most commonly cited algorithm in publications in Journal of Theoretical Biology and PLoS Computational Biology.

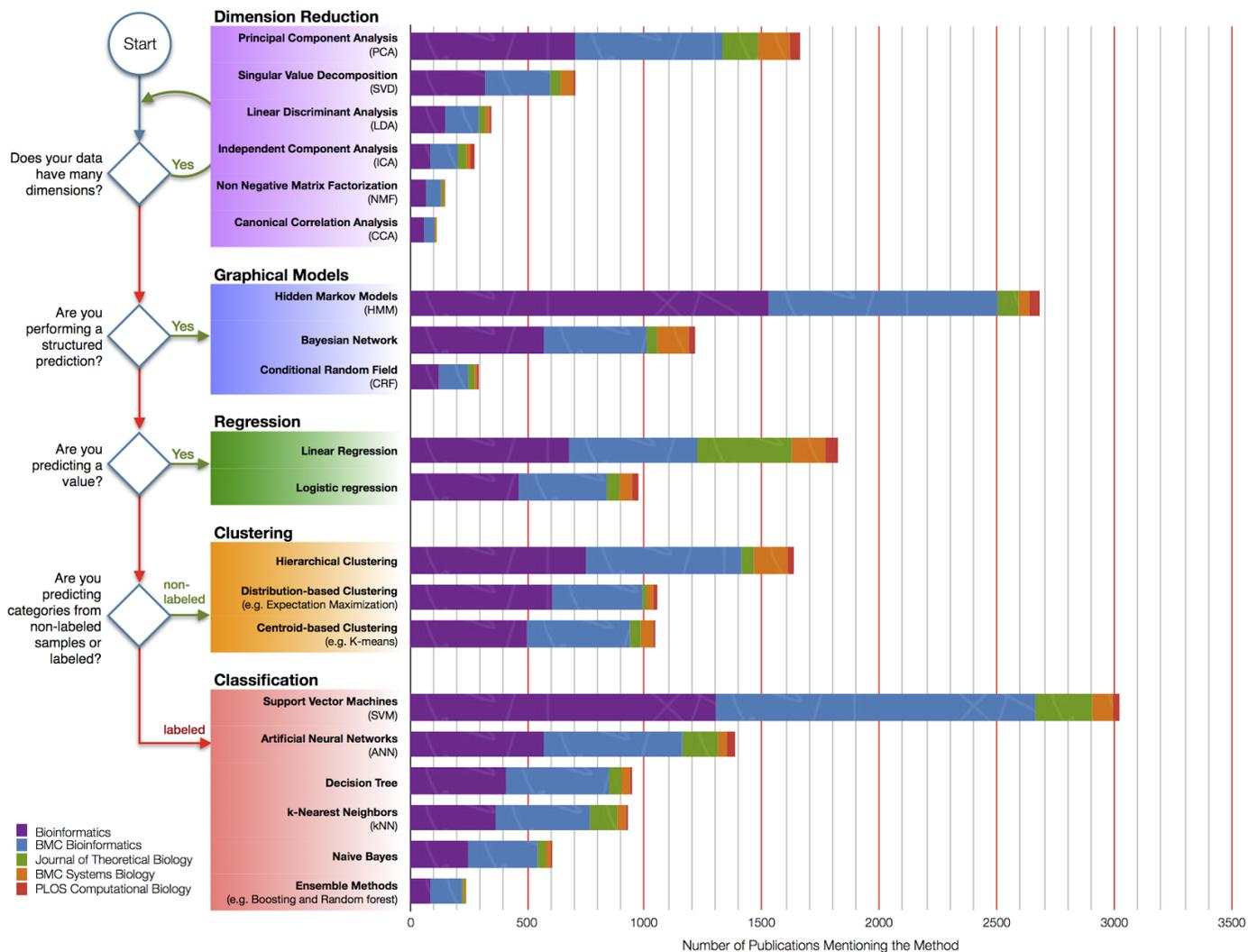

Figure 1 Machine Learning and Data Mining algorithms in the bioinformatics literature. The plot shows, the common process flow for choosing an algorithm along with the set of algorithms commonly used in each category. The values represent the number of publications which have mentioned the algorithm in top five bioinformatics journal venues.

## Current State of Scalable Machine Learning and Data Mining

The field of distributed systems has its origins in the late-1970's, when several researchers began investigating issues in multi computer systems. Prior to this time, research had concentrated on either single computer systems or parallel problem-solving systems that did not involve communication and synchronization among multiple computing components. Since this early research in distributed systems, the field has grown dramatically, with a wide variety of topics being addressed. In recent years, with distributed systems becoming more mature and reliable, application areas have been explored more extensively. Naturally, with the exponential rise of data, large-scale data analytics -- mining and learning from large quantities of data -- has been one such explored application areas. As a result, multiple distributed systems have been built to specifically address large-scale data analytics, among which few have become mature and usable enough to be utilized by non distributed systems experts. Apache

Mahout [12], Spark MLlib [13], and Google's TensorFlow [25] are the three notable ones, which rely on a large community of developers, trying to provide scalable alternatives to traditional single machine libraries/tools, such as R [16], Weka [17], and Octave [18]. Apache Mahout and Spark MLlib have been widely utilized in industry to analyze large amounts of data, such as implementing user interest models, advertisement targeting, and mining frequent pattern in emails [19].

Although these distributed tools have provided strong capabilities for industry, several important machine learning algorithms for bioinformatics from Figure 1 are missing (Table 1), such as Independent Component Analysis (ICA), Non Negative Matrix Factorization (NMF), Hidden Markov Models (HMM), Bayesian Network, k-Nearest Neighbors (kNN), and Artificial Neural Networks (ANN). This is partly due to the fact that distributed systems efforts have followed the most common MLDM textbook algorithms.

Table 1 Overview of large-scale tools for machine learning and data mining. Implemented algorithms are marked as "Yes", blank areas indicate "not implemented". Listed algorithms are the top machine learning and data mining techniques reflected in the Figure 1.

| Category | Algorithm | Apache Mahout [20] | Spark MLlib [21] | TensorFlow [25], CNTK [26] |
|---|---|---|---|---|
| Dimension Reduction | Principle Component Analysis (PCA) | Yes | Yes | |
| | Singular Value Decomposition (SVD) | Yes | Yes | |
| | Linear Discriminant Analysis (LDA) | | | |
| | Independent Component Analysis (ICA) | | | |
| | Non Negative Matrix Factorization (NMF) | | | |
| | Canonical Correlation Analysis (CCA) | | | |
| Graphical Models | Hidden Markov Models (HMM) | | | |
| | Bayesian Network | | | |
| | Conditional Random Field (CRF) | | | |
| Regression | Linear Regression | | Yes | |
| | Logistic regression | | Yes | |
| Clustering | Hierarchical Clustering | | Yes | |
| | Distributed-based Clustering (EM) | | Yes | |
| | Centroid-based Clustering (K-means) | Yes | Yes | |
| **Classification** | Support Vector Machines (SVM) | | Yes | |
| | Artificial Neural Networks (ANN) | | | Yes |
| | Decision Tree | | Yes | |
| | k-Nearest Neighbors (kNN) | | | |
| | Naive Bayes | Yes | Yes | |
| | Ensemble Methods (Boosting, Random forest) | Yes (random forest) | Yes (boosting and random forest) | |

## Challenges of Scalable Machine Learning and Data Mining

We next consider some of the challenges that have held back extensive development of scalable MLDM algorithms and infrastructures for use with big data. We begin by noting that the volume of data is not the only criterion for big data or the only challenge to scalability; there have been considerable discussions from both industry and academia on the definition, challenges and opportunities brought about by increased data as early as 2001 [22,23,24]. The characteristics and challenges of big data may be summarized by the "Vs model", i.e., Volume (data scale becomes increasingly big), Variety (various types and modalities of data), Velocity (rapid generation and timeliness of data), and Value (extract huge value from very low density). This definition is widely recognized since it highlights the meaning and necessity of big data, i.e., how to discover values from datasets with an enormous scale, various types, and rapid generation. As a large class of analytical algorithms, machine learning and data mining techniques are key to transforming big data into value and actionable knowledge in various domains including bioinformatics and the related fields of health care and precision medicine. To make this happen, the distributed systems researcher's has to design and implement *distributed*, *parallel*, *efficient*, *terminatable*, *provably correct* algorithms, and provide *usable* frameworks.

**Parallel, distributed, and efficient:** as the data grows so does the number of machines where the data resides. The effective analysis of this massive amount of data requires the MLDM methods to execute in a distributed manner and at an appropriate scale in a way that each machine can be assigned an independent task. While running these algorithms on a single machine often seems straightforward, the task of parallelizing and distributing them is exceedingly complicated. Many standard MLDM methods iteratively transform parameters during the learning process. For example, many classification and regression algorithms can be formulated as a convex optimization problem, i.e., the task of finding a minimizer of a convex function that depends on a variable vector. To solve the optimization problem, we iteratively refine a set of parameters until some termination condition is achieved. In parallel and distributed settings, in order to fully utilize the resources the systems experts are required to address traditional issues such as race conditions, deadlocks, distributed state, consistency, locking and communication protocols. Meanwhile, with the availability of Cloud computing services like Amazon EC2, Microsoft Azure, and Google Cloud Platform with on-demand access to large-scale computing and storage resources, the systems expert has to address further challenges. In the cloud environment, the computation time is costly, to reduce the time and cost, it is important to develop sophisticated schedulers and intelligently reduce the network overhead and I/O waste while utilizing CPU resources. Constructing a sophisticated resource manager and scheduler requires the careful reasoning about the system's architecture and complexities of parallel algorithm design. Failure is another critical issue in the cloud. Both software and hardware failure is the norm rather than the exception, hence the need for addressing issues such as fault-tolerance while simultaneously ensuring mathematical properties of the algorithms.

**Termination:** one of the mathematical properties of MLDM algorithms that has to be ensured is to design a termination condition that detects when a fixed point has been reached. The standard termination condition assessment for majority of MLDM techniques requires knowledge of the global state. In distributed setting, efficient termination assessment is challenging, usually resulting in a trade-off between computational performance and accuracy.

**Correctness:** the next challenge is to guarantee the consistency and convergence to a relevant result. In parallel and distributed settings, different computation nodes perform the execution of their task at different pace. Therefore, MLDM algorithms will have different convergence behaviour compared to their sequential version. In some cases, simultaneous (or misordered) execution can lead to race conditions, resulting in slow, unstable, and even corrupted convergence. Even though some MLDM algorithms are robust to approximate executions, choosing the right execution and consistency model has direct implications for the algorithm correctness.

**Usability:** to attain maximum performance, many parallel and distributed systems tools (e.g., MPI, PThreads) provide powerful and expressive primitives. Consequently, non-experts have a hard time of acquiring these tools. To achieve a level of adoption similar to R, Python, and Matlab, the systems expert has to design high-level abstractions, simplify the user's implementation, and insulate users from the complexities of consistency, races conditions, and deadlocks.

## On Deep Learning

In the last couple of years, new breakthroughs started happening in machine learning and artificial intelligence. Techniques started working much better, and new techniques have appeared, especially around artificial neural networks, and when they were applied to some long-standing and important use cases dramatically better results were gained [27]. These techniques have also shown to have much broader applications, especially in bioinformatics [28,29,30], clearly promising a fundamental change for the community. Notably, when we integrate multi-omics and clinical data to work on a path towards precision interventions, we will be able to answer healthcare and clinical questions better, and might answer questions that we could not really answer before at all. Hence, it is possible that the bioinformatics community like the tech industry will start recentering on these techniques.

However, while we have the algorithms, we do not have systems that are capable of providing the execution environment at large scale. Presently, we have a set of publicly available tools such as Microsoft's CNTK and Google's TensorFlow that are being developed but we are still far from the desired large scale. Not only do these algorithms run on very large data sets, but also, as is generally acknowledged, the more hidden layers and nodes in a neural network, the higher accuracy they can produce [31]. However, as the complexity of neural network increases so does the learning time. Consequently, the severely time and memory consuming process of complex neural network training and the existing big data challenges leaves the systems community with a fundamental challenge to tackle.

## The Long Road Ahead
Bioinformatics is a broad area that integrates techniques from several fields including data mining, machine learning, statistics, pattern recognition, and artificial intelligence for the analysis of data. Our survey identifies some of the most influential machine learning and data mining algorithms in the bioinformatics research community. We recognize that the approach is to an extent confounded by publications citing an algorithm, especially a popular algorithm, even though the publication itself does not use that algorithm. We also caution that this systematic review is not meant to serve as evidence of any general superiority of one algorithm over another.

In the era of big data there exist many research challenges for adapting and optimizing data-driven analytics. To overcome the data deluge in computation biology and bioinformatics [2], the focus has to be on collaboration between bioinformaticians, data scientists, and scalable computing experts to apply scalable algorithms, new storage and computation designs, and should aim to achieve the possibilities of large-scale data analysis with significant improvements in performance. This systematic review aims to guide computer scientists, especially those interested in scalable computing systems, to bioinformatics algorithms that merit their attention most, following the engineering maxim of "optimize the common case". We hope this paper can inspire more researchers in scalable computing to further explore these top algorithms, including their scalability and potential scalable alternatives.

## Acknowledgment
This research program is sponsored in part through RDA/US Data Share grant from the Alfred P. Sloan Foundation #G-2014-13746, and supported by grant 1U54GM114838 awarded by NIGMS through funds provided by the trans-NIH big data to Knowledge (BD2K) initiative (www.bd2k.nih.gov), and a collaborative award from the National Science Foundation (NSF award numbers CNS-1329686, CNS-1329737, CNS-1330142, and CNS-1330491), and in part by the Intramural Research Program of the National Institute on Aging, National Institutes of Health (Z01-AG000949-02). The views and conclusions in this document are those of the authors and should not be interpreted as necessarily representing the official policies, either expressed or implied, of NSF or NIH.

## Supplementary Note 1

Compiled list of well known machine learning and data mining algorithms. These top algorithms cover classification, clustering, regression, graphical model-based learning, and dimensionality reduction, which are among the most important topics in machine learning and data mining research and development. This list includes algorithms which were highly cited and algorithms that did not have at least 50 papers citing/using them, and were eliminated from the final graph:

**Decision Tree**: '"Decision Tree" OR "Decision Trees" OR "C4.5"'
**Ensembles**: '"Bootstrap aggregating" OR bootstrap bagging OR "AdaBoost" OR "LogitBoost" OR "Random Forests" OR "Random Forest"'
**SVM**: '"support vector machines" OR "support vector machine" OR SVM OR SVMs'
**Linear Regression**: '"Linear Regression" OR "Linear Regressions"'
**Naive Bayes**: '"Bayes Classifier" OR "Bayes Classifiers" OR "Naive Bayes"'
**Artificial Neural Networks**: '"Artificial Neural Networks" OR "Artificial Neural Network" OR "neural networks" training learning OR "neural network" training learning'
**Logistic regression**: '"Logistic Regression" OR "Logistic Regressions"'
**k-nearest neighbors**: 'KNN OR "k-NN" OR "k Nearest Neighbors" OR "k Nearest Neighbor" OR "k Nearest Neighbour" OR "k Nearest Neighbours"'
**Hierarchical Clustering**: '"Hierarchical Clustering"'
**Centroid-based clustering**: '"K-means" OR kmeans OR "k means" OR "Centroid-based clustering"'
**Distribution-based clustering:** '"Expectation Maximization" OR "Distribution-based clustering"'
**PCA:** '"Principal component analysis" OR "Principal components analysis" OR PCA'
**SVD:** '"singular value decomposition" OR "singular values decomposition" OR SVD'
**LDA:** '"Linear Discriminant Analysis" OR LDA'
**Canonical Correlation Analysis:** '"Canonical Correlation Analysis"'
**ICA:** '"Independent component analysis" OR "Independent components analysis"'
**NMF:** 'NMF OR "Non negative matrix factorization" OR "Non negative matrix factorizations"'
**Bayesian network:** '"Bayesian Networks" OR "Bayesian Network" OR "Bayes Model"'
**CRF:** '"Conditional random field" OR "Conditional random fields"'
**HMM:** 'HMM OR "hidden markov model" OR "hidden markov models"'

Eliminated algorithms which did not pass the 50 cited papers threshold:
**RVM:** '"Relevance vector machine" OR RVM'
**Density-based clustering:** '"Density-based clustering"'
**t-SNE**: '"t-distributed stochastic neighbor embedding" OR tSNE OR t-SNE'
**Local outlier factor:** '"Local outlier factor"'
**LDA_latent:** '"Latent Dirichlet allocation"'
**PageRank:** '"PageRank"'